\documentclass[useAMS,usenatbib]{mn2e}

\usepackage{graphicx}
\usepackage{natbib}

\title[The binary fraction in thirteen globular clusters]{The fraction of binary systems in the core of thirteen 
low-density Galactic globular clusters\thanks{Based
on ACS observations collected with the Hubble Space Telescope 
within the observing program GO 10755.}}
\author[Sollima et al.]{A. Sollima$^{1}$\thanks{E-mail:
antonio.sollima@bo.astro.it (AS)},  G. Beccari$^{2,3,4}$, F. R.
Ferraro$^{1}$, F. Fusi Pecci$^{2}$ and A. Sarajedini$^{5}$\\
$^{1}$Dipartimento di Astronomia, Universit\`a di Bologna, via Ranzani 1,
Bologna, 40127-I, Italy\\
$^{2}$INAF Osservatorio Astronomico di Bologna, via Ranzani 1,
Bologna, 40127-I, Italy\\
$^{3}$INAF Osservatorio Astronomico di Collurania, via M. Maggioni, 64100-I
Teramo, Italy\\
$^{4}$Dipartimento di Scienze della Comunicazione, Universit\`a di Teramo,
Teramo, 64100-I, Italy\\
$^{5}$Department of Astronomy, University of Florida, 32611 Gainesville, USA}
\begin{document}

\date{Accepted 2007 June 15; Received 2007 March 19; in original form
2007 July ??}

\pagerange{\pageref{firstpage}--\pageref{lastpage}} \pubyear{2007}

\maketitle

\label{firstpage}

\begin{abstract}
We used deep observations collected with ACS@HST to derive the 
fraction of binary systems in a sample of thirteen low-density Galactic globular 
clusters. 
By analysing the color distribution of Main Sequence stars we derived the
minimum fraction of binary systems required to reproduce the observed color-magnitude diagram
morphologies. We found that all the analysed globular clusters contain
a minimum binary fraction larger than 6\% within the core radius. 
The estimated global fractions of binary systems range from 10\% to 50\% depending on the cluster.
A dependence of the relative fraction of binary systems on the cluster 
age has been detected, suggesting that the binary disruption process within the cluster 
core is active and can significantly reduce the binary content in time.
\end{abstract}

\begin{keywords}
stellar dynamics -- methods: observational -- techniques: photometric -- 
binaries: general -- stars: Population II -- globular clusters: general
\end{keywords}

\section{Introduction}
Binary stars provide a unique
tool to determine crucial information about a variety of stellar
characteristics including mass, radius and luminosity. Fortunately, there are
ample opportunity to observe binary star systems: most stars are in fact in 
binary systems, at least in the solar
neighborhood (Duquennoy \& Mayor 1991). Since the first decades of the twentith century, the
study of binary star systems provided valuable information about the stellar
structure and evolution, such as the mass-luminosity and mass-radius relations
(Kuiper 1938; Huang \& Struve 1956).
Binarity, under particular conditions, induces the onset of nuclear reactions
leading to the formation of bright objects like novae and determines the fate 
of low-mass stars leading to SN~Ia explosions.

Binaries play also a key role in the dynamical evolution of stellar systems and stellar
populations studies. 
In collisional systems binaries provide the gravitational fuel that can delay
and eventually stop and reverse the process of core collapse in globular clusters (see
Hut et al. 1992 and references therein). 
Furthermore, the evolution of binaries in star clusters can produce peculiar stellar object of
astrophysic interest like blue stragglers, cataclysmic variables, low-mass X-ray
binaries, millisecond pulsars, etc. (see Bailyn 1995 and reference therein). 
The binary fraction is a key ingredient in chemical and dynamical models to 
study the evolution of galaxies and stellar systems in general.  

The main techniques used to derive the binary fraction in globular clusters are:
{\it i)} radial velocity variability surveys (Latham 1996; Albrow et al. 2001) {\it ii)} searches
for eclipsing binaries (Mateo 1996) and {\it iii)} searches for secondary
main-sequences (MS) in color-magnitude diagrams (CMD, Rubenstein \& Bailyn 1997).
The first two methods rely on the detection of individual binary systems in a given
range of periods and mass-ratios. The studies carried out in the past 
based on these methods argued for a deficiency of binary stars in globular
clusters compared to the field (Pryor et al. 1989; Hut 1992; Cote et al. 1996). However, the nature
of these two methods leads to intrinsic observational biases and a low detection 
efficiency.
Conversely, the estimate of the binary fraction on the basis of the analysis of
the number of stars displaced in the secondary MS represents a more efficient 
statistical approach and  
does not suffer of selection biases. In fact, any binary
system in a globular cluster is seen as a single star with a flux equal to the
sum of the fluxes of the two components. This effect 
locates any binary system sistematically at brighter magnitudes with respect to 
single MS stars, defining a secondary sequence in the CMD running parallel to 
the cluster MS that allows to distinguish them from other single MS stars. 
Until now, the binary fraction has
been estimated following this approach only in few globular clusters (Romani \& Weinberg 1991; 
Bolte 1992; Rubenstein \& Bailyn 1997; 
Bellazzini et al. 2002; Clark, Sandquist \& Bolte 2004; Zhao \& Bailyn 2005).

In this paper we present an estimate of the binary fraction in thirteen low-density 
Galactic globular clusters. We used the photometric survey carried out with
the {\it Advanced Camera for Surveys} (ACS) on board HST as a part of a Treasury
program (Sarajedini et al. 2007).  

In \S 2 we describe the observations, the data reduction techniques and the 
photometric calibration. In \S 3 the adopted method to determine the
fraction of binary systems is presented. In \S 4 we derived the minimum binary
fractions in our target globular clusters. \S 5 is devoted to the estimate of the
global binary fractions and to the comparison of
the measured relative fractions among the different globular clusters of our
sample. In \S 6 the radial distribution of binary systems is analysed. 
Finally, we summarize and discuss our results in \S 7.

\section{Observations and Data reduction}
\label{reduct}

The photometric data-set consists of a set of high-resolution images obtained
with the ACS on board HST through the F606W ($V_{606}$) and F814W ($I_{814}$) filters. 
The target clusters were selected on the basis of the following criteria:
\begin{itemize}
\item A high Galactic latitude ($b>15^{\circ}$) in order to limit the field
contamination;
\item A low reddening (E(B-V)$<$0.1) in order to avoid the occurrence of 
differential reddening;
\item A low apparent central density of stars\footnote{The apparent central density of 
stars has been calculated from the central surface density $\rho_{S,0}$ and the cluster distance d 
(from McLaughlin \& Van der Marel 2005) according to the following relation
$$\rho_{0}'=\rho_{S,0}d^2(\frac{2\pi}{21600})^2$$}
($log~\rho_{0}'<5~M_{\odot}~arcmin^{-2}$) in order to limit the effects of
crowding and blending.
\end{itemize}
Thirteen cluster passed these criteria namely NGC288, NGC4590, NGC5053,
NGC5466, NGC5897, NGC6101, NGC6362, NGC6723, NGC6981, M55, Arp 2, Terzan 7 and Palomar 12.
In Table 1 the main physical parameters of the above target clusters are listed.
The central density $\rho_{0}$, the core radii $r_{c}$ and the half-mass relaxation times $t_{r,r_{h}}$
 are from Djorgovski (1993), the age $t_{9}$  from Salaris \& Weiss (2002) and the global 
 metallicities $[M/H]$ from Ferraro et al. (1999)\footnote{For the clusters NGC6101, NGC6362, NGC6723 
 and Palomar 12 not included in the list of Ferraro et al. (1999) we transformed the metallicitiy 
 [Fe/H] from Zinn \& West (1984) into the global metallicity [M/H] following the prescriptions of 
 Ferraro et al. (1999).}.
Note that the analysed sample spans a wide range in age and metallicity containing 
only low-density ($log~\rho_{0}<2.75~M_{\odot}pc^{-3}$) globular clusters. 

\begin{figure}
 \includegraphics[width=8.7cm]{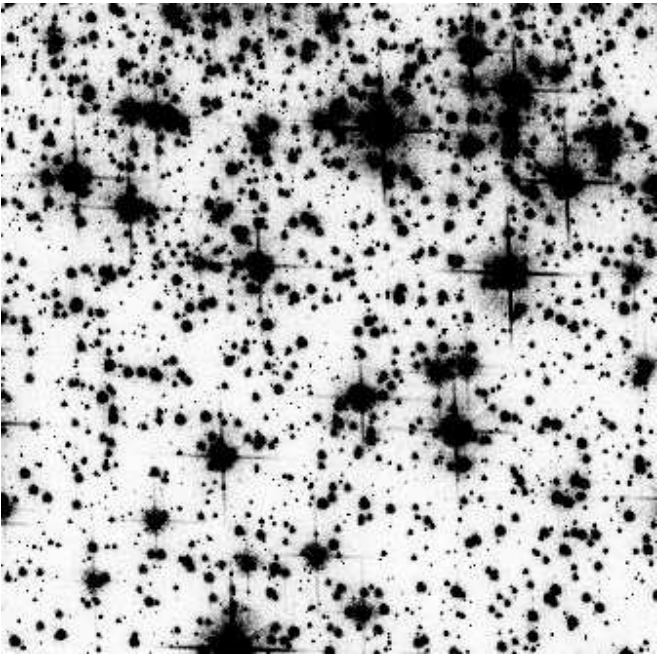}
\caption{Zoomed image of the central region of the globular cluster NGC6723, 
the most crowded cluster of our sample.}
\label{6723f}
\end{figure}

For each cluster the ACS field of view was centered on the cluster center.
We retrived all the available exposures from the ESO/ST-ECF Science Archive. 
The exposure times for each cluster in each filter are listed in Table 2.
All images were passed through the standard ACS/WFC reduction pipeline. 
Data reduction has been performed on the individual pre-reduced images
using the SExtractor photometric package (Bertin \& Arnouts 1996). 
The choice of the data-reduction software has been made after several trials using
the most popular PSF-fitting softwares. However, the shape of the PSF quickly 
varies along the ACS chip extension giving trouble to most PSF-fitting algorithms.
Conversely, given the small star density in these clusters, crowding does not affect the
aperture photometry, allowing to properly estimate the magnitude of stars. 
This is evident in Fig. \ref{6723f} where a zoomed portion of the central 
region of the cluster NGC6723 (the most crowded GC of our sample) is shown. 
Note that the surface density of stars in this field is $\leq 1.4~stars~arcsec^{-2}$.
For each star we measured 
the flux contained within a radius of 0.125" (corresponding to 2.5 pixels $\sim$ FWHM) from the star center. 
The source detection and the photometric analysis have been performed 
independently on each image. Only stars detected in three out four frames have been 
included in the final catalog. The most isolated and
brightest stars in the field have been used to link the aperture
magnitudes at 0.5" to the instrumental ones, after normalizing for exposure time. 
Instrumental magnitudes have been transformed into
the VEGAMAG system by using the photometric zero-points by Sirianni et al.
(2005).
Finally, each ACS pointing has been corrected for geometric distorsion using the
prescriptions by Hack \& Cox (2001).

Two globular clusters (NGC5053 and NGC5466) were already analyzed
by Sarajedini et al. (2007).
Our photometry has been compared with the photometric catalog already published
by these authors. 
The mean magnitude differences found are $\Delta V_{606}$ = -0.004 $\pm$ 0.012 
and $\Delta I_{814}$ = 0.004 $\pm$ 0.012
for NGC5053 and $\Delta V_{606}$ = -0.031 $\pm$ 0.012 and $\Delta I_{814}$ =
-0.020 $\pm$ 0.012 for NGC5466, which 
are consistent with a small systematic offset in both passbands.

\begin{figure*}
 \includegraphics[width=12.cm]{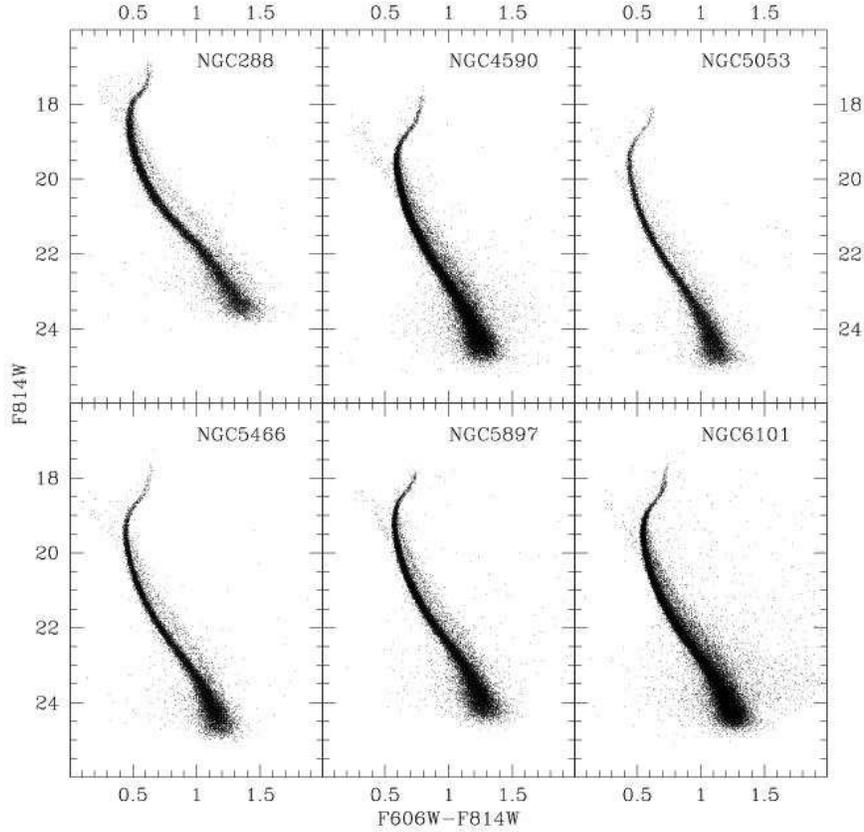}
\caption{$I_{814}, V_{606}-I_{814}$ CMDs of the target globular clusters NGC288,
NGC4590, NGC5053, NGC5466, NGC5897 and NGC6101.}
\label{cmd1}
\end{figure*}

\begin{figure*}
 \includegraphics[width=12.cm]{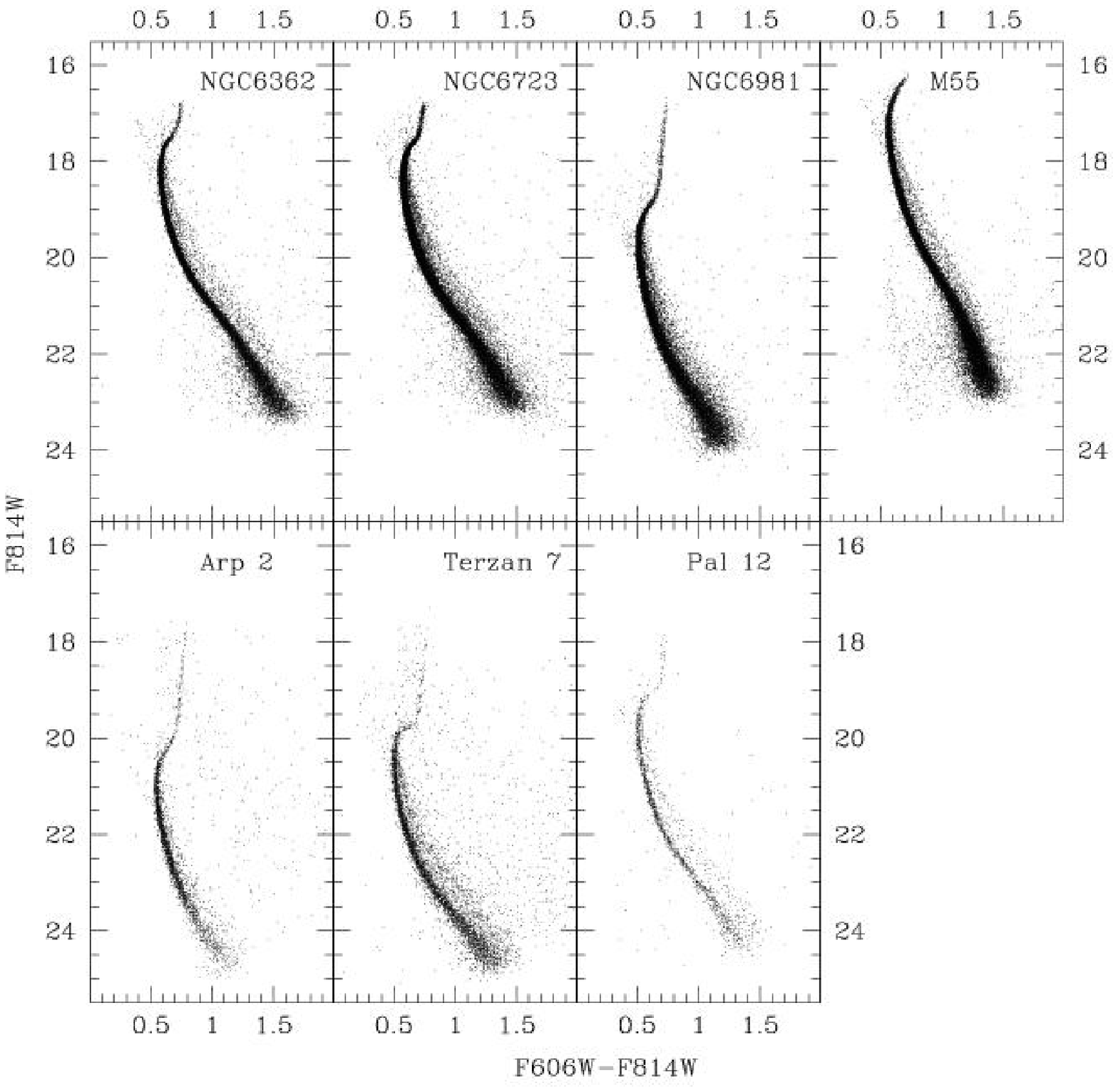}
\caption{$I_{814}, V_{606}-I_{814}$ CMDs of the target globular clusters NGC6362,
NGC723, NGC6981, Arp 2, M55, Terzan 7 and Palomar 12.}
\label{cmd2}
\end{figure*}

Fig. \ref{cmd1} and \ref{cmd2} show the ($I_{814}, V_{606}-I_{814}$) CMDs of the 
13 globular clusters in our sample. The CMDs sample the cluster
population from the sub-giant branch down to 5-6 magnitudes below the MS turn-off.
In all the target clusters the binary sequence is well defined and distinguishable 
from the cluster's MS. In the less
dense clusters (e.g. Terzan 7, Pal 12) binary stars appears to populate preferentially
a region of the CMD $\sim$0.752 mag
brighter than the cluster MS, approaching the equal-mass binary sequence (Eggleton, 
Mitton \& Whelan 1978).
In most clusters
a number of blue stragglers stars populating the bright part of the
CMD is also evident. 

\begin{table}
\label{riass}
 \centering
  \caption{Main physical parameters of the target globular clusters}
  \begin{tabular}{@{}lccccr@{}}
  \hline
   Name     & $log~\rho_0$           & $r_{c}$ & $t_{9}$ & $log~t_{r,r_{h}}$ & [M/H]\\
            & $M_{\odot}~pc^{-3}$    &   "     &   Gyr   &    Gyr            &        \\
 \hline
 NGC 288    & 1.80 &  85.20   &   11.3  &    8.99  & -0.85\\
 NGC 4590   & 2.52 &  41.35   &   11.2  &    8.90  & -1.81\\
 NGC 5053   & 0.51 &  134.40  &   10.8  &    9.59  & -2.31\\
 NGC 5466   & 0.68 &  116.50  &   12.2  &    9.37  & -1.94\\
 NGC 5897   & 1.32 &  118.70  &   12.3  &    9.31  & -1.44\\
 NGC 6101   & 1.57 &   69.25  &   10.7  &    9.22  & -1.40\\
 NGC 6362   & 2.23 &  79.15   &   11.0  &    8.83  & -0.72\\
 NGC 6723   & 2.71 &  56.81   &   11.6  &    8.94  & -0.73\\
 NGC 6981   & 2.26 &  32.09   &   9.5$^\ast$   &    8.93  & -1.10\\
 M55        & 2.12 &  170.8   &   12.3  &    8.89  & -1.41\\
 Arp 2      & -0.35&   96.03  &   7-11.5  &    9.46  & -1.44\\
 Terzan 7   & 1.97 &  36.51   &   7.4   &    9.03  & -0.52\\
 Palomar 12 & 0.68 &  65.83   &   6.4   &    9.03  & -0.76\\
 \hline
\end{tabular} 
$^\ast$ The age of NGC6981 has been taken from De Angeli et al. (2005)(see \S \ref{compar}). 
\end{table}

\begin{table}
\label{summ}
 \centering
  \caption{Observing logs}
  \begin{tabular}{@{}lccr@{}}
  \hline
   Name     & \# of exposures & Filter & Exposure time\\
            &                 &        &     (s)\\
 \hline
 NGC 288    & 4 & $V_{606}$ & 130\\
            & 4 & $I_{814}$ & 150\\
 NGC 4590   & 4 & $V_{606}$ & 130\\
            & 4 & $I_{814}$ & 150\\
 NGC 5053   & 5 & $V_{606}$ & 340\\
            & 5 & $I_{814}$ & 350\\
 NGC 5466   & 5 & $V_{606}$ & 340\\
            & 5 & $I_{814}$ & 350\\
 NGC 5897   & 4 & $V_{606}$ & 340\\
            & 3 & $I_{814}$ & 350\\
 NGC 6101   & 5 & $V_{606}$ & 370\\
            & 5 & $I_{814}$ & 380\\ 
 NGC 6362   & 4 & $V_{606}$ & 130\\
            & 4 & $I_{814}$ & 150\\ 
 NGC 6723   & 4 & $V_{606}$ & 140\\
            & 4 & $I_{814}$ & 150\\ 
 NGC 6981   & 4 & $V_{606}$ & 130\\
            & 4 & $I_{814}$ & 150\\ 
 M55        & 4 & $V_{606}$ & 70\\
            & 4 & $I_{814}$ & 80\\ 
 Arp 2      & 5 & $V_{606}$ & 345\\
            & 5 & $I_{814}$ & 345\\
 Terzan 7   & 5 & $V_{606}$ & 345\\
            & 5 & $I_{814}$ & 345\\		     
 Palomar 12 & 5 & $V_{606}$ & 340\\
            & 5 & $I_{814}$ & 340\\ 	  	   
 \hline
\end{tabular}
\end{table}

\section{Method}
\label{method}

As quoted in \S 1, any binary
system in a globular cluster is seen as a single star with a flux equal to the
sum of the fluxes of the two components. This effect produces a systematic 
overluminosity of these objects and a shift in color depending 
on the magnitudes of the two components in each passband. In a simple stellar
population the luminosity of a MS star is univocally connected with its mass.
In particular, stars with smaller masses have fainter magnitudes following a
mass-luminosity relation.  
So, named $M_{1}$ the mass of the most massive (primary) component in a given binary
system and $M_{2}$ the mass of the less massive (secondary) one, the magnitude of the 
binary system can be written as:
$$m_{sys}= -2.5~log(F_{M_{1}}+F_{M_{2}})+c$$
$$=m_{M_{1}}-2.5~log(1+\frac{F_{M_{2}}}{F_{M_{1}}})$$ 
In this formulation the shift in magnitude of the binary system can be viewed as 
the effect of the secondary star that perturbs the magnitude of the primary. 
The quantity $\frac{F_{M_{2}}}{F_{M_{1}}}$ depends on the mass ratio of the two
component ($q=\frac{M_{2}}{M_{1}}$). According to the definition of $M_{1}$ and
$M_{2}$ given above, the parameter $q$ is comprised in the range $0<q<1$. 
When q=1 (equal mass binary) the binary system will
appear $-2.5~log(2)\sim0.752$ mag brighter than the primary component.
Conversely, when $q$ approaches small values the ratio
$\frac{F_{M_{2}}}{F_{M_{1}}}$ becomes close to zero, producing a negligible
shift in magnitude with respect to the primary star.
Following these considerations, binary systems with small values of $q$ becomes
indistinguishable from MS stars when photometric errors are present.  
Hence, only binary systems with values of $q$ larger than a minimum value ($q_{min}$) 
are unmistakably distinguishable from single MS stars. 
For this reason, only a lower limit to the binary fraction can be directly derived
without assuming a specific distribution of mass-ratios $f(q)$.

In order to study the relative frequency of binary systems in our target
clusters we followed two different approaches:
\begin{itemize}
\item We derived the minimum number of binary systems by considering only the
fraction of binary systems with large mass-ratio values ($q>q_{min}$);
\item We estimated the global binary fraction by assuming a
given $f(q)$ and comparing the simulated CMDs with the observed ones.
\end{itemize}

A correct binary fraction estimation requires corrections for two important effects: {\it i)}
blended sources contamination and {\it ii)} field stars contamination.
In the following sections we describe the adopted procedure to take into account 
these effects.
 
\subsection{Blended sources}
\label{blend}

Chance superposition of two stars produces the same magnitude enhancement observed in
a binary system. For this reason it is impossible to
discern whether a given object is a physical binary or not. However, a
statistical estimate of the distribution of blended sources expected to populate the
CMD as a function of magnitude and color is possible by means of extensive artificial 
stars experiments (see Bellazzini et al. 2002).  

For each individual cluster the adopted procedure for the artificial star experiments has been
performed as follows:
\begin{itemize}
\item{The cluster mean ridge line has been calculated by averaging the colors of stars in the 
CMD over 0.2 mag boxes and applying a 2$\sigma$ clipping algorithm;}
\item{The magnitude of artificial stars has been randomly extracted from a
luminosity function (LF) modeled to 
reproduce the observed magnitude distribution of bright stars ($F814W < 22$) and to provide large 
numbers of faint 
stars down to below the detection limits of the observations ($F814W > 26$)\footnote{Note 
that the assumption for the fainter stars is only for statistical purposes, i.e. to 
simulate a large number of stars in the range of magnitude where significant losses due to 
incompleteness are expected.}. The color of each star has been obtained by deriving, for each 
extracted F814W magnitude, the corresponding F606W magnitude by interpolating on the 
cluster ridge line. Thus, all the artificial stars lie on the cluster ridge line in the CMD;}
\item{We divided the frames into grids of cells of known width (30 pixels) and randomly positioned 
only one artificial star per cell for each run\footnote{We constrain each artificial star to 
have a minimum distance (5 pixels) from the edges of the cell. In this way we can control the 
minimum distance between adjacent artificial stars. At each run the absolute position of the 
grid is randomly changed in a way that, after a large number of experiments, the stars are 
uniformly distributed in coordinates. Given the small stars density in the analysed cluster areas, 
the radial dependence of the completeness factor turns of to be neglegible.};}
\item{Artificial stars have been simulated using the Tiny Tim model of the ACS PSF 
(Krist 1995\footnote{The Tiny Tim version 6.3 updated to model the ACS PSF is 
available at http://www.stsci.edu/software/tinytim/}) 
and added on the original frames including Poisson photon noise. Each star has 
been added to both F606W and F814W frames. 
The measurement process has been repeated adopting the same procedure of the 
original measures and applying the same 
selection criteria described in Sect. \ref{reduct};}
\item{The results of each single set of simulations have been appended to a file until the 
desired total number of artificial stars has been reached. The final result for each subfield is a 
list containing the input and output values of positions and magnitudes.}
\end{itemize}

\begin{figure}
 \includegraphics[width=8.7cm]{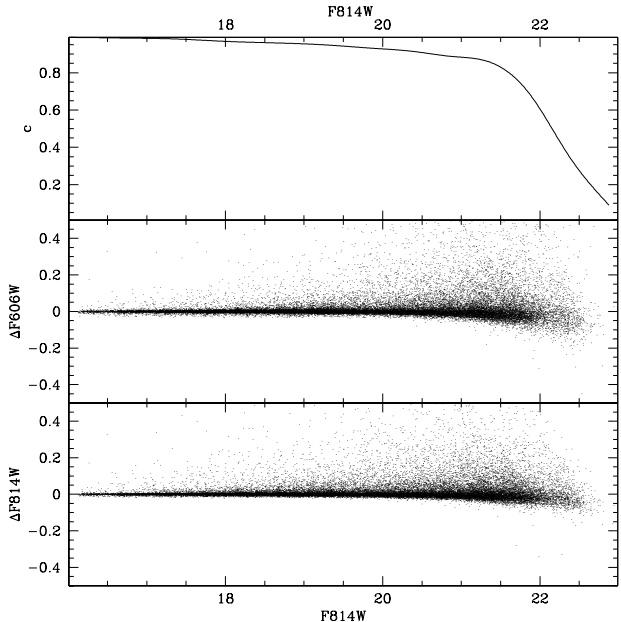}
\caption{Completeness factor $c$ as a function of the F814W magnitude ($upper~
panel$) for the target cluster M55. In $lower~panels$ the residuals between the 
input and output F606W and F814W magnitudes of artificial stars are shown.}
\label{compl}
\end{figure}

The residuals between the input and output $V_{606}$ and $I_{814}$ magnitudes
and the completeness factor as
a function of the $I_{814}$ magnitude are shown in Fig. \ref{compl} for the 
case of M55 as an example. As expected, the distributions of the magnitude residuals are not
symmetrical: a significant number of stars have been recovered with a brigther
output magnitude than that assigned in input. This effect is due to those stars
that blended with nearby real stars with similar (or larger) luminosity.  
More than 100,000 artificial stars have been produced for each cluster 
providing a robust estimate of the blending
contamination together with the levels of
photometric accuracy and completeness in all the regions of the CMD and
throughout the cluster extension. 

\subsection{Field stars}
\label{field}

Another potentially important contamination effect is due to the presence of background and foreground field 
stars that contaminate the binary region of the CMD.
To account for this effect, we used the Galaxy model of Robin et al. (2003). A
catalog covering an area of 0.5 square degree around each cluster center (from Djorgovski \&
Meylan 1993) has been retrived.
A sub-sample of stars has been randomly extracted from the entire catalog scaled to the
ACS field of view ($202"\times~202"$).      
The V and I Johnson-Cousin magnitudes were converted into the ACS photometric system by means of the
transformations of Sirianni et al. (2005).
For each synthetic field star, a star with similar input magnitude ($\Delta I_{814}<0.1$) 
has been randomly extracted from the artificial stars catalog. If the artificial
star has been recovered in the output catalog the $V_{606}$
and $I_{814}$ magnitude shifts with respect to its input magnitudes have been 
added. This procedure accounts for the effects of incompleteness, photometric errors and blending. 

\section{The minimum Binary fraction}
\label{minfrac}

As pointed out in \S \ref{method} there is a limited range of mass-ratio values ($q>q_{min}$) where it is 
possible to clearly distinguish binary systems from single MS stars. 
The value of $q_{min}$ depends on the photometric accuracy 
(i.e. the signal-to-noise S/N ratio) of the data. 
The approach presented in this section allows to estimate the fraction of binaries with $q>q_{min}$ 
that represents a lower limit to the global cluster binary fraction.

\begin{figure}
 \includegraphics[width=8.7cm]{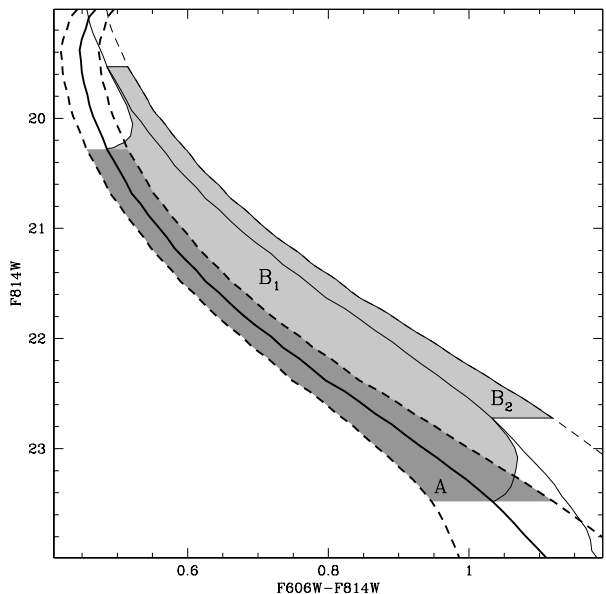}
\caption{Selection boxes used to select the $MS sample$ (dark grey area) and the 
$binary~sample$ (grey area). The solid
thick line marks the MS mean ridge line, the solid thin line marks the equal-mass
binary sequence, dashed lines mark the $4~\sigma$ range used to define the
selection boxes $A, B_{1}$ and $B_{2}$ (see \S \ref{minfrac}).}
\label{box}
\end{figure}

In the following we will refer to the binary fraction $\xi$ as the ratio between 
the number of binary systems whose primary star has a mass comprised in a 
given mass range ($N_{b}$) and the number of cluster members in the same mass
range ($N_{tot}=N_{MS}+N_{b}$)\footnote{This quantity can be easily converted in
the fraction $\xi$' of $stars~in~binary~systems$ ($N_{b,s}$) with respect to the cluster stars
($N_{tot,s}$) considering that $N_{b,s}=2~N_{b}$ according to the relation 
$$\xi'=\frac{2\xi}{1+\xi}$$}.
To derive an accurate estimate of this quantity we adopted the following procedure:

\begin{enumerate}
\item We defined an $I_{814}$ magnitude range that extends from 1 to 4 magnitudes below the 
cluster turn-off. In this magnitude range the completeness factor is always $\phi>50\%$;
\item We converted the extremes of the adopted magnitude range ($I_{up}$ and $I_{down}$) into masses 
($M_{up}$ and $M_{down}$) using the mass-luminosity
relation of Baraffe et al. (1997). To do this, the V and I Johnson-Cousin magnitudes of the Baraffe et al. (1997)
models were converted into the ACS photometric system by means of the transformations by Sirianni et
al. (2005). 
For our target clusters we assumed the metallicities listed by Ferraro et al. (1999), the distance moduli 
and reddening coefficients listed by Harris (1996) and the extinction coefficients $A_{F814W}=2.809~E(B-V)$ and
$A_{F814W}=1.825~E(B-V)$ (Sirianni et al. 2005). Small shifts in the distance
moduli ($\Delta (m-M)_{0}<0.1$) have been applied in order to match the overall
MS-TO shape; 
\item We defined three regions of the CMD (see Fig. \ref{box}) as follows:
\begin{itemize}
\item A region (A) containing all stars with $I_{down}<I_{814}<I_{up}$ and a 
color difference from the MS mean ridge line
smaller then 4 times the photometric error corresponding to their magnitude
(dark grey area in Fig. \ref{box}). This area contains all the single MS stars in the above magnitude 
range and binary systems with $q<q_{min}$;
\item We calculated the location in the CMD of a binary system formed by a primary star of mass
$M_{up}$ (and $M_{down}$ respectively) and different mass-ratios $q$ ranging from 0 to 1. 
These two tracks connect the MS mean ridge line with the equal mass binary 
sequence (which is 0.752 mag
brighter than the MS ridge line) defining an area ($B_{1}$) in the CMD. 
This area contains all the binary systems with $q<1$ and whose primary component 
has a mass $M_{down}<M_{1}<M_{up}$;
\item A region ($B_{2}$) containing all stars with magnitude
$I_{down}-0.752<I_{814}<I_{up}-0.752$ and whose color difference from the 
equal mass binary sequence 
is comprised between zero and 4 times the photometric error corresponding to 
their magnitude. This area is populated by binary systems with $q\sim1$ that are shifted to the 
red side of the equal-mass binary sequences because of photometric errors;
\end{itemize}
\item We considered single MS stars all stars contained in A ($MS~sample$), binary 
stars all stars contained in $B_{1}$ and
$B_{2}$ but not in A ($binary~sample$, grey area in Fig. \ref{box});
\item Since the selection boxes defined above cover two different 
regions of the CMD with different
completeness levels, we assigned to each star lying in the $MS~sample$ and in
the $binary~sample$ a completeness factor $c_{i}$ according to its magnitude (Bailyn et al. 1992). 
Then, the corrected number of stars in each sample 
($N_{MS}^{CMD}$ and $N_{bin}^{CMD}$) has been calculated as
$$N=\sum_{i} \frac{1}{c_{i}}$$
\item We repeated steps {\it (iv)} and {\it (v)} for the samples of artificial 
stars and field stars,
obtaining the quantities $N_{MS}^{art}$ and $N_{bin}^{art}$ for the 
$artificial~stars~sample$ and
$N_{MS}^{field}$ and $N_{bin}^{field}$ for the $field~stars~sample$;
\item We calculated the normalization factor $\eta$ for the 
$artificial~stars~sample$ by comparing the
number of stars in the MS selection box
$$\eta=\frac{N_{MS}^{CMD}}{N_{MS}^{art}}$$
\item The minimum binary fraction, corrected for field stars and blended sources, turns out to be
$$
\xi_{min}=\frac{N_{bin}^{CMD}-N_{bin}^{field}-\eta~N_{bin}^{art}}{(N_{MS}^{CMD}-N_{MS}^{field})+(N_{bin}^{CMD}-N_{bin}^{field}-\eta~N_{bin}^{art})}$$
\end{enumerate}

Since the target clusters in our sample are located at different distances, the
ACS field of view covers different fractions of the cluster's extent.
The procedure described above has been conducted considering only cluster stars (and
artificial stars) located inside 
one core radius ($r_{c}$, adopted from Djorgovski 1993). 

The obtained minimum binary fractions $\xi_{min}$ for the clusters in our sample are listed in
Table 3. The typical error (calculated
by taking into account of the Poisson statistic and the uncertainties in the 
completeness corrections) is of the order of 1\%. As can be noted, the minimum binary fraction $\xi_{min}$ is 
larger than 6\%  
in all the clusters of our sample. Therefore, this value seems to represent a lower limit to
the binary fraction at least in low-density ($log~\rho_0 < 2.75 M_{\odot} pc^{-3}$, see Table 1) 
globular clusters.

\section{The global binary fraction}
\label{true}

The procedure described above allowed us to estimate the minimum 
binary fraction $\xi_{min}$ without any (arbitrary) 
assumption on the distribution of mass-ratios  $f(q)$. However, caution must be used when 
comparing the derived binary fraction among the different clusters of our sample. 
In fact, the definition of the $MS~sample$ and $binary~sample$ given in \S \ref{minfrac} depends on the 
photometric accuracy (e.g. the S/N ratio) that vary from cluster to cluster.
An alternative approach consists in the simulation of a binary population which 
follows a given distribution $f(q)$ and in the comparison between the 
color distribution of simulated stars and the observed CMD.
Until now there are neither theoretical
arguments nor observational constraints to the shape of $f(q)$ in globular
clusters.
Studies on binary systems located in the local field suggest
that the overall shape of $f(q)$ can be reproduced by extracting randomly 
secondary stars from the observed Initial Mass Function (IMF, Tout 1991).
Fisher et al. (2005) estimated the mass-ratio distribution $f(q)$ in the binary 
population of the local field (at distances $d<100~pc$).
They found that most binary
systems are formed by similar mass components ($q\sim 1$). 

\begin{figure}
 \includegraphics[width=8.7cm]{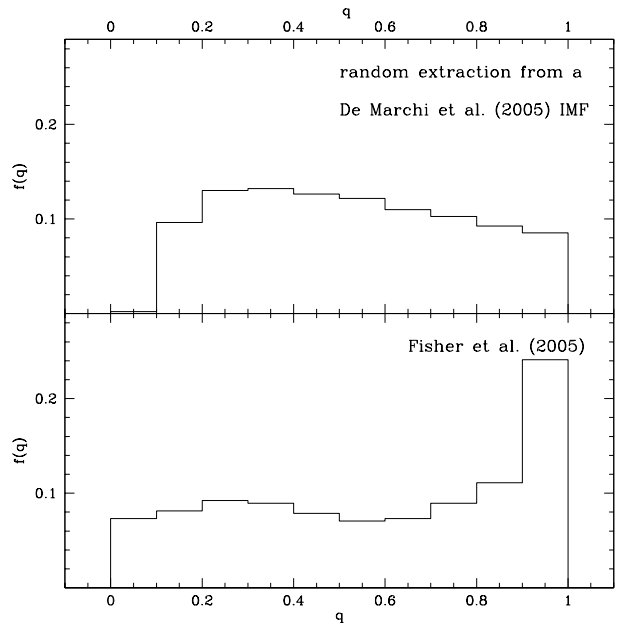}
\caption{Distribution of mass-ratios of 100,000 binary stars simulated in the
magnitude range $I_{down}<I_{814}<I_{up}$ from random extractions from a De
Marchi et al. (2005) IMF ($upper~panel$). The distribution of mass-ratios
adopted from Fisher et al. (2005) is shown in the $bottom~panel$.}
\label{fq}
\end{figure}

In the following we calculate the binary fraction $\xi$ in the target clusters
assuming two different shape of $f(q)$: {\it i)} a distribution
constructed by extracting random pairs of stars from the De Marchi et al. (2005) IMF 
(see Fig. \ref{fq} upper panel) and {\it ii)} the distribution $f(q)$ measured by
Fisher et al. (2005, see Fig. \ref{fq} lower panel).

\subsection{$\xi_{RA}:~f(q)$ from random associations}
\label{ran}

In the case of binary stars formed by random associations between stars of
different masses the general scheme adopted for an assumed binary fraction $\xi$ 
has been the following:
\begin{enumerate}
\item Artificial star $I_{814}$ magnitudes have been
converted into masses by means of the mass-luminosity relation of Baraffe et al.
(1997). Then, a number of $N~(1-\xi)$ artificial stars were
extracted from a De Marchi et al. (2005) IMF, where N is the number of stars in
the observed catalog. This sample of stars reproduces the MS population of each
cluster taking into account also of blended sources;
\item The binary population has been simulated as follows:
\begin{description}
\item {\it a)} A number of $N'(>\xi~N)$ pairs of stars were extracted randomly from a De
Marchi et al. (2005) IMF;
\item {\it b)} The $V_{606}$ and $I_{814}$ magnitudes of the two
components were derived adopting the mass-luminosity relations of Baraffe et al.
(1997) and the corresponding fluxes were summed in order to obtain the $V_{606}$ 
and $I_{814}$ magnitudes of the unresolved binary system;
\item {\it c)} For each binary system, a star with similar input magnitude ($\Delta I_{814}<0.1$) 
has been randomly extracted from the artificial stars catalog. If the artificial
star has been recovered in the output catalog the $V_{606}$
and $I_{814}$ magnitude shifts with respect to its input magnitudes have been 
added. This procedure accounts for the effects of incompleteness, photometric errors and
blending; 
\item {\it d)} The final binary population has been simulated
by extracting a number of $\xi N$ objects from the entire catalog.
\end{description}  
\item The field stars catalog (obtained as described in \S \ref{field}) was 
added to the simulated sample;
\item The ratio between the number of objects lying in the selection boxes 
defined in \S \ref{minfrac} ($r_{sim}=\frac{N_{bin}^{sim}}{N_{MS}^{sim}}$) has 
been calculated and compared to that measured in
the observed CMD ($r_{CMD}=\frac{N_{bin}^{CMD}}{N_{MS}^{CMD}}$); 
\item Steps from {\it (i)} to {\it (iv)} have been repeated 100 times and a
penalty function has been calculated as
$$\chi^2=\sum_{i=1}^{100}~(r_{sim_{i}}-r_{CMD})^{2}$$ 
\end{enumerate}
The whole procedure has been repeated for a wide grid of binary 
fractions $\xi$ and a probability distribution as a function of $\xi$ has been
produced.
The value of $\xi$ which minimizes the penalty function $\chi^2$ has been 
adopted as the most probable. 
The error on the estimated binary fraction has been estimated by estimating the 
interval where the $\chi^2$ account for the 68.2\% probability ($\sim 1 \sigma$)
to recover the measured quantity.

\begin{figure*}
 \includegraphics[width=12cm]{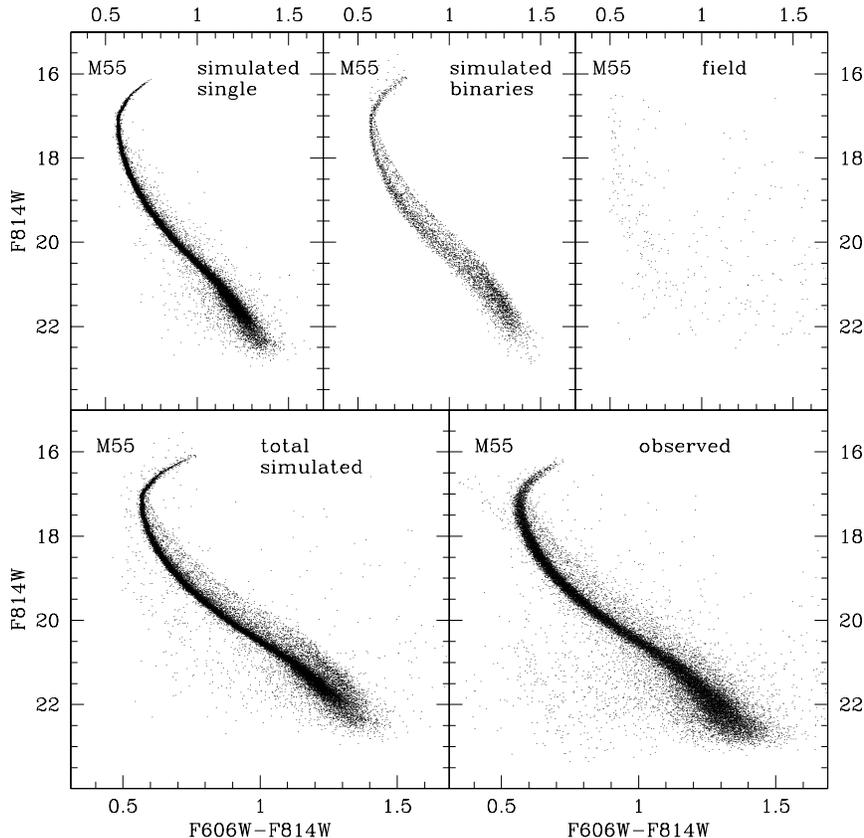}
\caption{Simulated ($lower~left~panel$) and observed ($lower~right~panel$) CMD 
of M55. In the $upper~panels$ the individual CMDs of the simulated single stars
 ($upper~left~panel$), binaries ($upper~central~panel$) and field stars
($upper~right~panel$) are shown.}
\label{simul}
\end{figure*}

\begin{figure}
 \includegraphics[width=8.7cm]{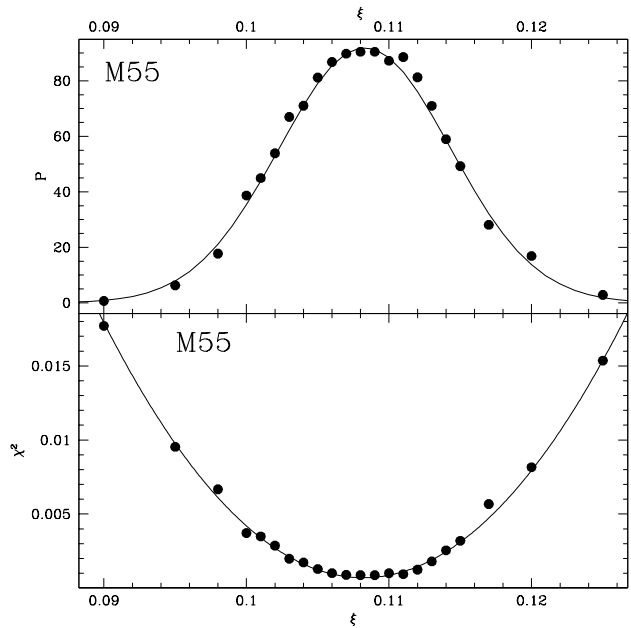}
\caption{Distribution of the calculated $\chi^2$ as a function of the assumed 
binary fraction for M55 ($bottom~panel$). A parabolic fit to the data is showed.
In the $top~panel$ the associated probability as a function of the assumed 
binary fraction is shown.}
\label{chi}
\end{figure}

A typical iteration of the procedure described above is showed in 
Fig. \ref{simul} where a simulated CMD of M55 is compared with the observed 
one. 
In Fig. \ref{chi} the distribution of $\chi^2$ and the related probability 
as a function of the assumed value of $\xi$ is shown.

The global binary fractions $\xi_{RA}$ for the target clusters 
are listed in Table 3. As can be noted, most of the analysed clusters harbour
a binary fractions $10\%<\xi<20\%$ with the exceptions of four cluster
(NGC6981, Arp 2, Terzan 7 and Palomar 12) which show a
significantly larger binary fraction ($\xi>35\%$).
We want to stress that, although this method is independent on the S/N ratio and
allows to derive the global binary fraction taking into account also of the hidden 
binary systems with $q<q_{min}$, it is subject to a number of systematic uncertainties
essentially due to unknown distribution of binary mass-ratios.
In fact, the binary fractions derived following the technique described above 
have a strong dependence on the low-mass end of the IMF whose
exact shape is still debated (see Kroupa 2002 and references therein). 
In particular, an increase of the fraction of low-mass stars significantly increases 
the probability to obtain binaries with low-mass secondaries (i.e. with small 
mass ratios $q$). This effect would produce a 
significant overestimate of the binary fraction.

The mass-ratios distribution derived from the above procedure, computed for a 
population of 100,000 binaries with $I_{down}<I_{814}<I_{up}$, is 
shown in Fig. \ref{fq} (upper panel). 
This distribution significantly differs from that observed by 
Halbwachs et al. (2003) and Fisher et al. (2005) (but see also Duquennoy \& Mayor 1991). 
In particular, most binary stars present small values of $q$
($q<0.5$) which produce a large number of hidden binaries. Thus, the 
binary fractions estimated in the observed clusters following this approach are probably 
sistematically overestimated. In the following we refer to this estimate as
$\xi_{RA}$ assuming it as a reasonable upper limit to the global binary fraction.

\subsection{$\xi_{F}:~f(q)$ from Fisher et al. (2005)}

As an alternative choice, we assumed a distribution of mass-ratios $f(q)$
similar to that derived by Fisher et al. (2005) from observations of
spectroscopic binaries in the solar neighborood (at distances $d<100~pc$).
The adopted mass-ratios distribution $f(q)$ is shown in Fig. \ref{fq} 
(lower panel).
Although this distribution is subject to significant observational 
uncertainties and is derived for binary systems in a different environment, 
it represents one of the few observational constraints to $f(q)$ which can 
be found in literature.

The adopted procedure to derive the binary fraction $\xi$ is the same 
described above but for the simulated binary population (point {\it (ii)a}). In
this case in fact, a number of $N'(>\xi~N)$ mass-ratios were extracted from the
distribution $f(q)$ shown in Fig. \ref{fq} (lower panel). Then, for each of the
$N'$ binary systems the mass of the 
primary component has been extracted from a 
De Marchi et al. (2005) IMF and the mass of the secondary component has been 
calculated. All the other steps of the procedure remain unchanged.

The calculated binary fractions $\xi_{F}$ are listed in Table 3. As expected, the
values of $\xi_{F}$ estimated following the assumption of a Fisher et al. (2005) 
$f(q)$ are comprised between the minimum binary fraction $\xi_{min}$ 
and the binary fraction estimated by random associations $\xi_{RA}$. Note
that neither the ranking nor the relative proportions of the binary fractions estimated
among the different clusters of the sample appear to depend on
the assumption of the shape of $f(q)$.  

For some clusters of our sample the binary fraction were already 
estimated in previous works. Bellazzini et al. (2002) and Bolte (1992) 
estimated a binary fraction comprised in the range $10\%<\xi<20\%$ for NGC288 by adopting a technique similar to the one
adopted here. These estimates are in good agreement with the result obtained in the
present analysis ($\xi\sim12\%$). Yan \& Cohen (1996) measured a binary fraction of $21\%<\xi<29\%$
in NGC5053 on the basis of a radial velocity survey. Our estimate suggests a
slightly smaller binary fraction in this cluster ($\xi\sim 11\%$). Note that 
the estimate by Yan \& Cohen (1996) is based on the detection of 6 binary
systems in a survey of 66 cluster members in a limited range of periods and
mass-ratios. The uncertainty of this approach due to the small statistic is
$\sim 10\%$ and can account for the difference between their estimate and the 
one obtained in the present analysis. 

In the following section we compare the obtained binary fractions among the clusters of
our sample as a function of their physical parameters. 

\begin{table}
\label{fractab}
 \centering
  \caption{Binary fractions estimated for the target globular clusters}
  \begin{tabular}{@{}lcccr@{}}
  \hline
   Name     & $\xi_{min}$ & $\xi_{F}$ & $\xi_{RA}$ & $\sigma_{\xi}$\\
            &    \%       &  \%       &   \%       &  \% \\
 \hline
 NGC 288    & 6  & 11.6 & 14.5 & 1.0 \\
 NGC 4590   & 9  & 14.2 & 18.6 & 2.5 \\
 NGC 5053   & 8  & 11.0 & 12.5 & 0.9 \\
 NGC 5466   & 8  &  9.5 & 11.7 & 0.7 \\
 NGC 5897   & 7  & 13.2 & 17.1 & 0.8 \\
 NGC 6101   & 9  & 15.6 & 21.0 & 1.3 \\
 NGC 6362   & 6  & 11.8 & 12.7 & 0.8 \\
 NGC 6723   & 6  & 16.1 & 21.8 & 2.0 \\
 NGC 6981   & 10 & 28.1 & 39.9 & 1.6 \\
 M55        & 6  &  9.6 & 10.8 & 0.6 \\
 Arp 2      & 8  & 32.9 & 52.1 & 3.6 \\
 Terzan 7   & 21 & 50.9 & 64.9 & 2.9 \\
 Palomar 12 & 18 & 40.8 & 50.6 & 6.6 \\
 \hline
\end{tabular}
\end{table}

\subsection{Cluster to cluster comparison} 
\label{compar}

Our sample contains thirteen low-density Galactic globular clusters spanning a large range of
metallicity, age and structural parameters (see Table 1).
We used the results obtained in the previous section to compare the core binary
fraction $\xi$ among the clusters of our sample as a function of their
main general and structural parameters in order to study the efficiency of the
different processes of formation and destruction of binary systems.

We correlated the core binary fraction derived according to the different
assumptions described in the previous sections ($\xi_{min}, \xi_{F}$ and
$\xi_{RA}$) with the cluster's ages
($t_{9}$, from Salaris \& Weiss 2002), global metallicity ([M/H], from Ferraro 
et al. 1999)
central density 
($\rho_{0}$) and half-mass relaxation time ($t_{r,r_{h}}$, from Djorgovski 1993), destruction rate 
($\nu$, from Gnedin \& Ostriker 1997) and different structural parameters (mass $M$, concentration $c$, 
binding energy $E_{b}$, half-mass radius $r_{F}$, mass-luminosity ratio $M/L$, 
velocity dispersion $\sigma_{v}$ and escape velocity $v_{e}$) adopted from 
McLaughlin \& Van der Marel (2005). 
Of course, most of the quantities listed above are correlated. 

The ages of two clusters, namely Arp2 and NGC6981, need a comment. 
According to Salaris \& Weiss (2002), the age of Arp 2 is comparable to those of the oldest Galactic 
globular clusters ($t_{9}\sim11.3$). The same conclusion has been reached by 
Layden \& Sarajedini (2000).
Conversely, Buonanno et al. (1995) and Richer et al. (1996) 
classified it as a young globular cluster with an age comparable within 1 Gyr to 
those of Terzan 7 and Palomar 12.
Given the debated question on the age of this cluster we excluded it from the following analysis.  
The globular cluster NGC6981 is not included in the list of Salaris \& Weiss (2002). 
An estimate of the 
age of this globular cluster has been presented by De Angeli et al. (2005).
We converted the ages measured by De Angeli et al. (2005) 
into the Salaris \& Weiss (2002) scale.   
Hence we adopted for this cluster an age of 9.5 Gyr.

In order to estimate the degree of dependence of $\xi$ on the different
clusters parameters we applied the $Bayesian~Information~Criterion$ test (Schwarz 1978) to our dataset.
We assumed the binary fraction $\xi$ as a
linear combination of a subsample of $p$ parameters ($\lambda_{i}$) selected among those listed
above. 
$$\xi_{f}=\alpha_{p+1}+\sum_{i=1}^{p} \alpha_{i}\lambda_{i}$$
Given a value of $p$, for any choice of the $p$ parameters we best-fit
our dataset with the above relation and calculated the quantity
$$BIC=\ell_{p}-\frac{p}{2}~log~N$$
where $\ell_{p}$ is the logarithmic likelihood calculated as
$$\ell_{p}=log ~L_{p} =  \sum_{j=1}^{N} log ~Pr_{j,p}$$
$$ = \sum_{j=1}^{N} log~(\frac{e^{-\frac{(\xi_{j}-\xi_{f,j})^{2}}{2\sigma_{\xi}^{2}}}}{\sigma_{\xi}\sqrt{2\pi}})$$
Where N is the dimension of our sample (N=13) and $\sigma_{\xi}$ is the residual
of the fit.
The $p$ parameters that maximize the quantity BIC are the most probable
correlators with $\xi$.
The above analysis gives the maximum value of BIC for $p$=1 and
$\lambda_{p}=t_{9}$. All the higher-order correlations appears as
non-significant.

The same result has been obtained considering all the three estimates of $\xi$.
A Spearman-rank correlation test gives probabilities $>99\%$ that the 
variables $\xi$ and $t_{9}$ are correlated, for all the considered estimates of
$\xi$.

\begin{figure}
 \includegraphics[width=8.7cm]{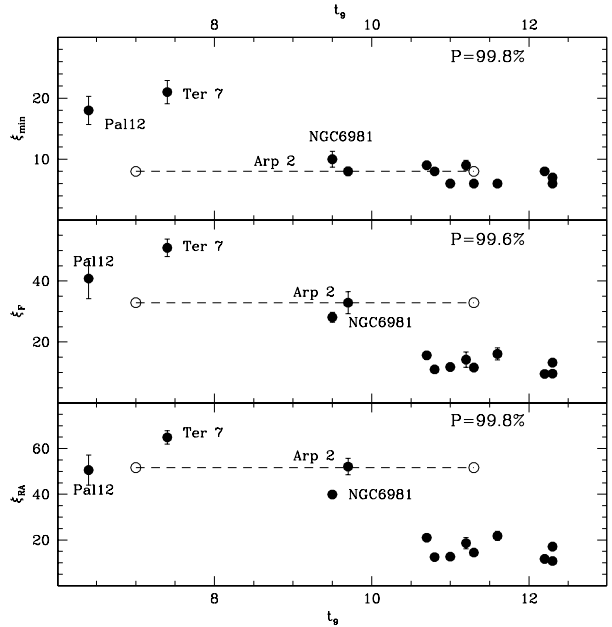}
\caption{Minimum ($upper~panel$), estimated ($middle~panel$) and maximum ($lower~panel$) 
binary fractions as a function of cluster age for the target clusters 
in our sample. For the cluster Arp 2 the upper and lower limit are marked as open points.}
\label{corrage}
\end{figure}

In Fig. \ref{corrage} the core binary fractions $\xi_{min}, \xi_{F}$ and
$\xi_{RA}$ are plotted as a function of the clusters age. 
All the clusters of our sample that present a large core binary fraction
($\xi_{F}>25\%$) are sistematically younger than the other clusters. 

Given the large systematic uncertainties involved in the estimate of the global
binary fraction the above result can be considered only in a 
qualitative sense. However, the above analysis indicates that the age seems to
be the
dominant parameter that determines the binary fraction in globular clusters
belonging to this structural class.

\section{Binaries radial distribution}
\label{radial}

\begin{figure}
 \includegraphics[width=8.7cm]{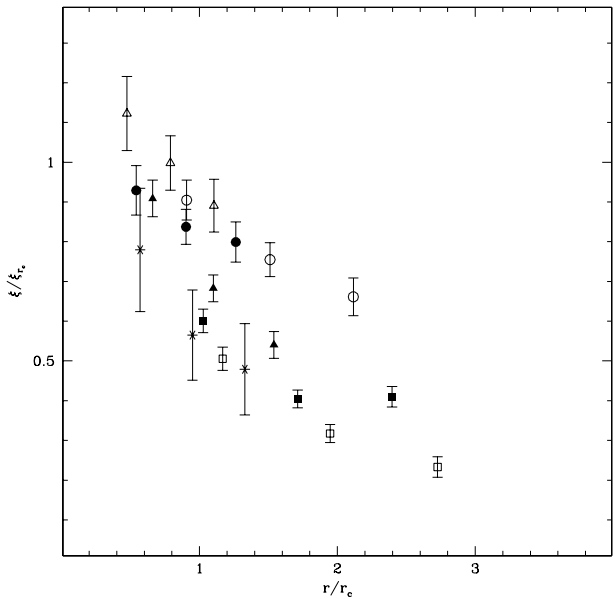}
\caption{Binary fraction (in unit of the core binary fraction) as a function of the
distance from the cluster center (in unit of core radii) for the target 
clusters NGC4590 (open circles), NGC6101 (filled circles), 
NGC6362 (open triangles), NGC6723 (filled triangles), NGC6981 (open squares), Terzan 7 (filled squares) 
and Palomar 12 (asterisks).}
\label{rad}
\end{figure}

Being bound systems, binary stars dynamically behave like a single star with a 
mass equal to the sum of the masses of the two components. After a time-scale comparable to the cluster 
relaxation time, binary systems have smaller mean velocities than single less massive stars, populating preferentially the most
internal regions of the cluster.
Since all the
globular clusters in our sample have a central relaxation time shorter than their
age, binary stars are expected to be more centrally concentrated with respect to
the other cluster stars.
In order to test this hypothesis we calculated for each target cluster the 
binary fraction $\xi$ (following the procedure described in \S \ref{minfrac}) 
in three annuli of 500 pixels width located at three different distances from the cluster center.
We noted that in seven (out of thirteen) globular clusters of our sample (namely
NGC4590, NGC6101, NGC6362, NGC6723, NGC6981, Terzan 7 and Palomar 12) 
there is evidence of radial segregation of binary systems toward the cluster center. 
In Fig. \ref{rad} the binary fractions (in unit of core fraction $\xi$) measured at different distances 
from the cluster's centers in these seven clusters are shown.  
The binary fraction decreases by a factor 2 at two core radii with respect 
to the core binary fraction.
A Kolmogorov-Smirnov test made on the $MS~sample$ and $binary~sample$ (as defined
in \S \ref{minfrac}) yields for these clusters probabilities smaller than 0.05\% that
the two samples are drawn from the same distribution.  
Note that in most clusters the radial segregation of binary systems is visible
also within the core radius, indicating that mass segregation is
a very efficient process in these clusters.
In the other six clusters the small number of stars and/or the small radial
coverage do not allow to detect a significant difference in the radial
distribution of binary stars.

\section{Discussion}

In this paper we analysed the binary population of thirteen low density 
Galactic globular clusters with the aim of studying their frequency and distribution.  

In all the analysed globular clusters the minimum binary fraction contained 
within one core radius is greater than 6\%. This quantity seems to represent a 
lower limit to the binary fraction in globular clusters of this structural 
class. This lower limit poses a firm constraint to the efficiency of the 
mechanism of binaries disruption. The existing estimates of the binary fraction
in low-density globular clusters (Yan \& Mateo 1994; Yan \& Reid 1996; Yan \& Cohen 1996) 
agree with this lower
limit. On the other hand, in high-density clusters the present day binary fraction 
appears to be smaller ($<4-9\%$ see Cool \& Bolton 2002 and Romani \& Weinberg 1991 for 
the case of NGC6397, M92 respectively) as 
expected because of the increasing efficiency of the 
disruption through close encounters and of stellar evolution (Ivanova et al.
2005). 
According to the theoretical simulations of Ivanova et al. (2005) the
present day binary fraction in a stellar system with a small central density
($10^{3} M_{\odot}pc^{-3}$) should be $<30\%$ of its initial fraction.
Following these considerations the initial binary fraction in our target 
globular clusters could be $>20-60\%$, comparable to that observed in the solar 
neighborhood (Abt \& Levy 1976; Duquennoy \& Mayor 1991; Reid \& Gizis 1997).

The comparison between the estimated relative binary fractions among the
clusters of our sample suggests that the age is
the dominant parameter that determines the fraction of surviving binary systems.
This result can be interpreted as an indication that the disruption of soft 
binary systems through close 
encounters with other single and/or binary stars is still efficient in low
density globular clusters also in the last 5 Gyr of evolution. 
Unfortunately, there are no estimates of the binary fraction in globular
clusters younger than 6 Gyr to test the efficiency of the process of binary
disruption in the early stages of evolution.
Note however that estimates of the binary fraction in open clusters (with ages
$<3~Gyr$) gives values as high as 30-50\% (Bica \& Bonatto 2005). 

The comparison between the radial distribution of binary systems with respect 
to MS stars indicates that binary systems are more concentrated toward 
the central region of most of the clusters of our sample. This evidence, already
found in other past works (Yan \& Reid 1996; Rubenstein \& Baylin 1997; 
Albrow et al. 2001; Bellazzini et al. 2002; Zhao \& Baylin 2005) 
is the result of the kinetic energy equipartition that lead binary systems to
settle in the deepest region of the cluster potential well. 
   
\section*{acknowledgements}
This research was supported by contract ASI-INAF I/023/05/0 and PRIN-INAF 2006. 
We warmly thank Michele Bellazzini and the anonymous referee for their helpful 
comments and suggestions and 
Paolo Montegriffo for assistance during catalogs cross-correlation.

\label{lastpage}


\begin{thebibliography}{99}

\bibitem[Abt et al.(1976)]{AL76} Abt H. A., Levy S. G., 1976, ApJS, 30, 273
\bibitem[Albrow et al.(2001)]{A01} Albrow M. D., Gilliland R. L., Brown T. M.,
Edmonds P. D., Guhathakurta P., Sarajedini A., 2001, ApJ, 559, 1060
\bibitem[Baraffe et al.(1997)]{B97} Baraffe I., Chabrier G., Allard F., Hauschildt P. H., 
1997, A\&A, 327, 1054
\bibitem[Bailyn et al. (1992)]{B92} Bailyn C. D., Sarajedini A., Cohn H., Lugger P. M., 
Grindlay J. E., AJ, 103, 1564
\bibitem[Baylin (1995)]{B95} Bailyn C. D., 1995, ARA\&A, 33, 133
\bibitem[Bellazzini et al. (2002)]{B02} Bellazzini M., Fusi Pecci F., Messineo
M., Monaco L., Rood R. T., 2002, AJ, 123, 509
\bibitem[Bertin et al.(1996)]{BA96} Bertin E., Arnouts S., 1996, A\&AS, 117, 393
\bibitem[Bica et al.(2005)]{BB05} Bica E., Bonatto C., 2005, A\&A, 431, 943
\bibitem[Bolte (1992)]{Bo92} Bolte C. D., 1992, ApJS, 82, 145
\bibitem[Buonanno et al. (1995)]{Bu95} Buonanno R., Corsi C. E., Fusi Pecci F., 
Richer H. B., Fahlman G. G., 1995, AJ, 109, 650
2004, AJ, 138, 3019
\bibitem[Clark et al. (2004)]{CSB04} Clark L. L., Sandquist E. L., Bolte M.,
2004, AJ, 138, 3019
\bibitem[Cool et al. (2002)]{CB02} Cool A. M., Bolton A. S., 2002, in
"Stellar Collisions, Mergers and their Consequences",  M. M. Shara eds., San
Francisco ASP
Conf. Ser., 263, 163
\bibitem[Cote et al. (1996)]{C96} Cote P., Pryor C., McClure R. D., Fletcher J.
M., Hesser J. E., 1996, AJ, 112, 574
\bibitem[De Marchi et al. (2005)]{D05} De Marchi G., Paresce F., Portegies Zwart
 S., 2005, in
"The Initial Mass Function 50 years later", E. Corbelli and F. Palle eds., 
Springer Dordrecht, 327, 77
\bibitem[Djorgovski (1993)]{D93} Djorgovski S., 1993, in "Structure and Dynamics of 
Globular Clusters", S. Djorgovski and G. Meylan eds., ASP Conf. Ser., 50, 373
\bibitem[Duquennoy et al.(1991)]{D91} Duquennoy A., Mayor M., 1991, A\&A, 248,
485
\bibitem[Eggleton et al. (1978)]{EMW78} Eggleton P., Mitton S., Whelan J.,
1978, ApL, 19, 101
\bibitem[Ferraro et al. (1999)]{F99} Ferraro F.R., Messineo M., Fusi Pecci F.
, De Palo M. A., Straniero O., Chieffi A., Limongi M., 1999, AJ, 118, 1738
\bibitem[Fisher et al. (2005)]{F05} Fisher J., Schroder K.P., Smith R.C., 2005, 
MNRAS, 361, 495
\bibitem[Gnedin et al.(1997)]{GO97} Gnedin O. Y., Ostriker J. P., 1997, ApJ, 474,
223
\bibitem[Hack et al.(2001)]{HC01} Hack W., Cox C., 2001, Instrum. Sci. Report,
2001-008
\bibitem[Halbwachs et al.(2003)]{H03} Halbwachs J. L., Mayor M., Udry S., Arenou
F., 2003, A\&A, 397, 159
\bibitem[Harris (1996)]{H96} Harris W. E., 1996, AJ, 112, 1487
\bibitem[Huang et al.(1956)]{HS56} Huang S. S., Struve O., 1956, AJ,
61, 300
\bibitem[Hut et al. (1992)]{H92} Hut P. et al., 1992, PASP, 104, 981
\bibitem[Ivanova et al. (2005)]{I05} Ivanova N., Belczynski K., Fregeau J. M.,
Rasio F. A., 2005, MNRAS, 358, 572
\bibitem[Krist (1995)]{K95} Krist J., 1995, in "Astronomical Data Analysis
Software and Systems IV", R. A. Shaw, H. E. Payne \& J. J. E. Hayes eds., ASP
Conf. Ser., 77, 349
\bibitem[Kuiper (1938)]{K38} Kuiper G. P., 1938, ApJ, 88, 472
\bibitem[Kroupa (2002)]{K02} Kroupa P., 2002, Sci, 295, 82
\bibitem[Latham (1996)]{L96} Latham D. W., 1996 in "The Origins, Evolution and
Destinies of Binary Stars in Clusters", E. F. Milone \& J. C. Mermilliod eds.,
San Francisco ASP Conf. Ser., 90, 31
\bibitem[Layden et al. (2000)]{LS00} Layden A., Sarajedini A., 2000, AJ, 119, 1760
\bibitem[Mateo (1996)]{M96} Mateo M., 1996 in "The Origins, Evolution and
Destinies of Binary Stars in Clusters", E. F. Milone \& J. C. Mermilliod eds.,
San Francisco ASP Conf. Ser., 90, 21
\bibitem[Mc Laughlin et al. (2005)]{MV05} McLaughlin D. E., Van der Marel R. P.,
2005, ApJS, 161, 304
\bibitem[Pryor (1989)]{P89} Pryor C., McClure R. D., Fletcher J. M., Hesser J.
E., 1989 in "Dynamics of Dense Stellar Systems", D. Merritt eds.,
Cambridge Univ. Press, p. 175
\bibitem[Reid et al. (1997)]{RG97} Reid I. N., Gizis J. E., 1997, AJ, 113, 2246
\bibitem[Richer et al.(1996)]{R96} Richer H. B., Harris W. E., Fahlman G. G., 
Bell R. A., Bond H. E., Hesser J. E., 1996, ApJ, 463, 602
\bibitem[Robin et al.(2003)]{R03} Robin A. C., Reil\'e C.,  Derri\`ere S.,
Picaud S., 2003, A\&A, 409, 523
\bibitem[Romani et al. (1991)]{RW91} Romani R. W., Weinberg M. D., 1991, ApJ, 372, 487
\bibitem[Rubenstein et al. (1997)]{RB97} Rubenstein E. P., Bailyn C. D., 1997,
ApJ, 474, 701
\bibitem[Salaris et al. (2002)]{SW02} Salaris M., Weiss A., 2002, A\&A, 388, 492
\bibitem[Sarajedini et al. (2007)]{S07} Sarajedini A. et al., 2007, ApJ, in
press, astro-ph/0612598 
\bibitem[Schwarz (1978)]{S78} Schwarz G., 1978, Annals of Statistics, 6, 461
\bibitem[Sirianni et al.(2005)]{S05} Sirianni et al., 2005, PASP, 117, 1049
\bibitem[Tout (1991)]{T91} Tout C. A., 1991, MNRAS, 250, 701
\bibitem[Yan et al. (1994)]{YM94} Yan L., Mateo M., 1994, AJ, 108, 1810
\bibitem[Yan et al. (1996)]{YC96} Yan L., Cohen J. G., 1996, AJ, 112, 1489
\bibitem[Yan et al. (1996b)]{YR96} Yan L., Reid M., 1996, MNRAS, 279, 751
\bibitem[Zhao et al. (2005)]{ZB05} Zhao B., Bailyn C. D., 2005, AJ, 129, 1934
\bibitem[Zinn et al. (1984)]{ZW84} Zinn R., West M. J., 1984, ApJS, 55, 45
\end{thebibliography}
\end{document}